\documentclass[12pt]{iopart}
\usepackage{amssymb}
\usepackage[dvips]{color,graphicx}
\usepackage{latexsym}

\begin{document}
\title{New Formulation of Causal Dissipative Hydrodynamics: Shock wave
propagation}
\author{Ph. Mota, G. S. Denicol, T. Koide and T. Kodama}

\begin{abstract}
The first 3D calculation of shock wave propagation in a homogeneous QGP has
been performed within the new formulation of relativistic dissipative
hydrodynamics which preserves the causality. We found that the relaxation
time plays an important role and also affects the angle of Mach cone.
\end{abstract}

\address{Instituto de F\'{\i}sica, 
Universidade Federal do Rio de Janeiro, C.P. 
68528, 21945-970 Rio de Janeiro, RJ, Brazil}

\section{Introduction}

One of the most important questions to be clarified in the hydrodynamical
approach to the relativistic heavy ion physics is the effect of dissipative
processes. The second order theory, first proposed by Muller and developed
by Israel and Stewart, has been considered standard approach for this
problem \cite{IS}. But it is quite complex and involves many unknown
parameters from the point of QCD dynamics so that its complete application
to practical problems such as relativistic heavy ion reactions has not been
done yet \cite{2}.

In this work, we propose an alternative approach to this question \cite{3}.
We show that the physical origin of the second order theories can easily be
understood in terms of memory effects. The irreversible current modified by
the memory effects becomes consistently with causality and sum rules \cite{4}%
. Based on this idea, we introduced the memory effect to the relativistic
dissipative hydrodynamic of Landau \cite{5}, where we introduce only one
extra parameter, the relaxation time $\tau _{relax},$ in addition to the
usual viscosity coefficients of the Navier-Stokes equation. The resulting
equation becomes hyperbolic \cite{3}.

The effect of viscosity is also important when we discuss the possible
generation and propagation of shock waves in the QCD medium created in the
process of relativistic heavy ion collisions. As discussed extensively in
this conference\cite{QM2006}, it has been suggested that a high energy jet
propagating in the QGP may generate a Mach cone and observables associated
with such phenomena may bring important information of the genuine
hydrodynamical properties of the matter \cite{6}. The dynamical simulation
of shock wave generation is very difficult even for the non-relativistic
regime. A full 3D simulation of shock wave dynamics has never been done for
the heavy ion collisions. In this work, we apply our formulation to the
calculation of full 3D relativistic (causal) shock wave problem. The
implementation of our method to the existing ideal hydro-codes is
straightforward, particularly to those based on the local Lagrangian
coordinate system such as SPheRIO \cite{7}. We organize the present work as
follows. In the next section, we briefly introduce our formalism and discuss
its application to the generation of shock waves. In Section 3, we present
some results of 3D calculation of shock wave propagation within the causal
formulation of the dissipative hydrodynamics. In Section 4, we discuss the
result and perspectives.

\section{Causal Approach to the Relativistic Dissipative Hydrodynamics}

The fundamental problem of the first order theory like the Navier-Stokes
theory comes from the fact that the diffusion equation is parabolic. The
physical origin of this problem can be followed up to the fact that the
irreversible current $\vec{j}$ is assumed to be proportional to a
thermodynamic force $\vec{F}$ as 
\begin{equation}
\vec{j}=-L\vec{F},  \label{Curr}
\end{equation}%
where the Onsager coefficient $L$ is, in general, a function of
thermodynamic quantities. Usually $\vec{F}$ is related to the inhomogeneity
in the density. When the microscopic rearrangement time scale is not
negligible compared to the time scale of the change in the irreversible
current, then the above should be replaced by the equation of motion for the
current, 
\begin{equation}
\frac{\partial \vec{j}}{\partial t}=-\frac{1}{\tau _{R}}L\vec{F}\left(
t\right) -\frac{1}{\tau _{R}}\vec{j}.  \label{Irrev}
\end{equation}%
where $\tau _{R}$ is the relaxation time. For very small $\tau _{R},$ we
recover Eq.(\ref{Curr}). Thus Eq.(\ref{Curr}) can be understood as the large
viscous limit of the damped motion, where the velocity (current) is
proportional to the force (Aristotelian vision). It can be shown that the
above modification is enough to convert the parabolic nature of a diffusion
equation to hyperbolic one \cite{3}.

For the relativistic hydrodynamics, we have to consider several different
kind of thermodynamical forces related to the velocity and density
inhomogeneity. They are $F=\partial _{\alpha }u^{\alpha },\ \ \ F_{\mu
}=\partial _{\mu }\alpha $ and$\ F_{\mu \nu }=\partial _{\mu }u_{\nu },$%
where $u^{\mu }$ is the four-velocity of the fluid, $\alpha =\mu /T$ with $%
\mu $ the chemical potential. These inhomgeneities generate the
corresponding irreversible currents and the analogous equations to Eq.(\ref%
{Irrev}) for them should be 
\begin{equation}
\tau _{R}\frac{d\Pi }{d\tau }=-\zeta \partial _{\alpha }u^{\alpha }-\Pi ,\ \
\tau _{R}\frac{d\widetilde{\pi }^{\mu \nu }}{d\tau }=\eta \partial ^{\mu
}u^{\nu }-\widetilde{\pi }^{\mu \nu },\ \ \tau _{R}\frac{d\widetilde{\nu }%
^{\mu }}{d\tau }=-\kappa \partial ^{\mu }\alpha -\widetilde{\nu }^{\mu },
\label{heat}
\end{equation}%
where $d/d\tau =u^{\mu }\partial _{\mu },$ and $\zeta ,$ $\eta $ and $\kappa 
$ are bulk viscosity, shear viscosity and thermal conductivity coefficients,
respectively. The energy-momentum tensor is expressed with these currents as 
$T^{\mu \nu }=\varepsilon u^{\mu }u^{\nu }-\left( p+\Pi \right) P^{\mu \nu
}+P^{\mu \nu \alpha \beta }\widetilde{\pi }_{\alpha \beta },$ where $P^{\mu
\alpha }=g^{\mu \nu }-u^{\mu }u^{\nu }$ and $P^{\mu \alpha \nu \beta }$ is
the double symmetric traceless projection, $P^{\mu \nu \alpha \beta }=\frac{1%
}{2}\left( P^{\mu \alpha }P^{\nu \beta }+P^{\mu \beta }P^{\nu \alpha
}\right) -P^{\mu \nu }P^{\alpha \beta }/3,$ whereas the conserved baryon
number current is given by $N^{\mu }=nu^{\mu }+P^{\mu \alpha }\widetilde{\nu 
}_{\alpha }.$ The hydrodynamic equations are 
\begin{equation}
\partial _{\mu }T^{\mu \nu }=0,\ \ \partial _{\mu }N^{\mu }=0.  \label{hydro}
\end{equation}

\section{Shock wave}

Although Eqs.(\ref{heat},\ref{hydro}) together with the equation of state
give the complete description of the hydrodynamical motion of the system, in
practice, some additional care to be taken, especially for the simulation of
the shock wave dynamics. Whenever there exists a shock wave, always occurs
an entropy production through the shock front. In an idealized
hydrodynamical approach, the shock front is a discontinuity in
thermodynamical quantities in a hydrodynamic solution. Mathematically
speaking, it should be treated as the boundary condition to connect two
distinct hydrodynamic solutions. Physically, it is not a real discontinuity,
but a quick change of the density in the region where the local equilibrium
is not satisfied. Thus it has a finite thickness at least a few times of the
mean-free path (typical microscopic scale of distance) for a stationary
shock. Under a dynamical condition such as relativistic heavy ion
collisions, the compression shock may have much more larger thickness due to
the many complicated local transient properties.

To reproduce true shock wave phenomena, the full degrees of freedom of the
hydrodynamics, together with a proper boundary condition correctly
connecting to regions through the non-equilibrated domain of the shock, are
required. The usual numerical approach of hydrodynamics excludes such a
possibility from the beginning. Since there exist no short wavelength
excitation modes due to the finite discretization, the energy and momentum
conservation required by the hydrodynamics result in a very rapidly
oscillations of the variables near the shock region. These quick
oscillations are somewhat compensating the thermal energy associated to the
entropy production throughout the shock front. In order to avoid these
oscillations, von Neumann and Richtmeyer\cite{Neuman} introduced the method
of pseudo-viscosity. The idea is to put the bulk viscosity in the shock
region to replace the entropy production through the shock front. The form
of the pseudo-viscosity is chosen so that asymptotic values of thermodynamic
quantities connects smoothly from one side to the other through the shock,
satisfying the Huguniot-Rankine boundary condition. For the relativistic
extension, we have to transform this formalism in the causal form. Thus, we
introduce the causal pseudo-viscosity by using the bulk viscosity
coefficient $\zeta $ in Eq.(\ref{heat}) as $\zeta =-\zeta _{0}s\left[
1-\alpha \theta \right] $ for $\theta <0$ and $0$ otherwise where $\theta
=\partial _{\mu }u^{\mu }$ and $\zeta _{0}$ and $\alpha $ are constants
proportional to the space resolution scale of the numerical solution. From
the condition of causality, the relaxation time associated to this
pseudo-viscosity should satisfy $\tau _{R}>\frac{3\zeta }{2\left(
p+\varepsilon \right) }.$This means, the larger the viscosity, the larger
the relaxation time. 
\begin{figure}[tbp]
\includegraphics[scale=0.6]{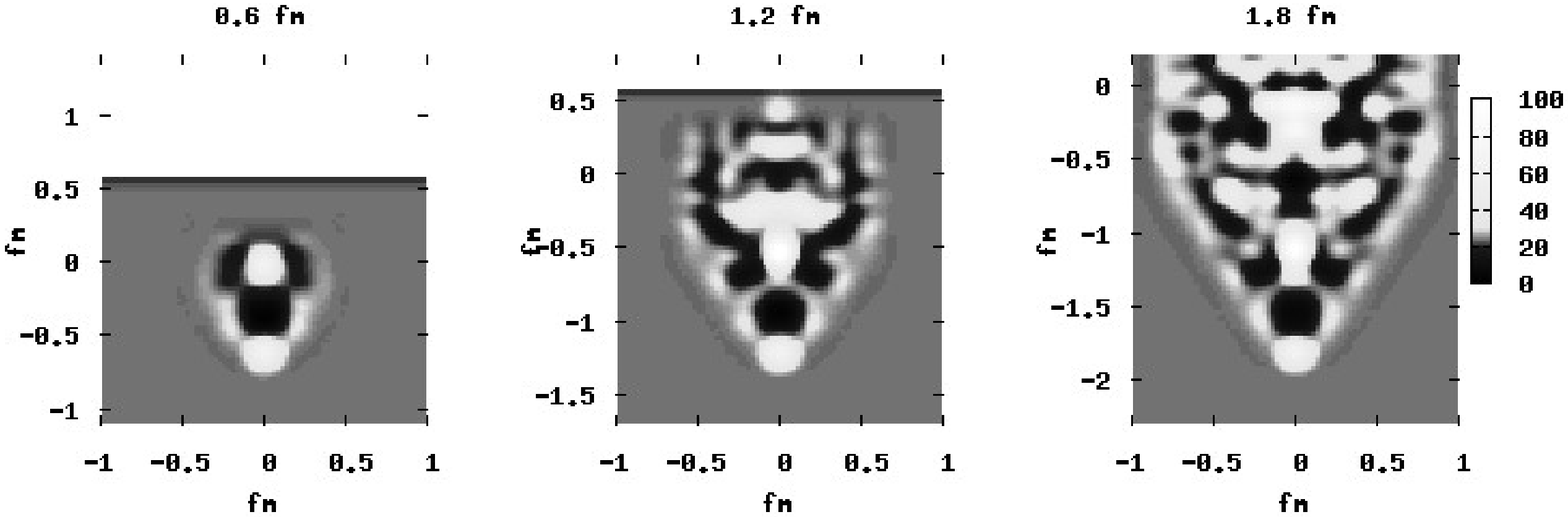}
\end{figure}
In Fig.1, we show the example of creation and propagation of shock wave
induced by a small QGP drop injected into the homogeneous QGP fluid at a
relativistic energy ($\gamma =10$ ). Here, we take a massless ideal gas
equation of state, and the incident drop has half the temperature of the
background in the local rest frame. The relaxation time necessary to
maintain the causality in this calculation is found to be $5fm/c$. This
seems to be large, but the effect of shock wave requires a large viscosity
so that a large relaxation time is necessary. We can see the generation of
Mach cone and also the turbulent structure after the shock cone. We found
that the cone angle depends on the viscosity and also the related relaxation
time $\tau _{R}.$ For a finite $\tau _{R}$ the Mach cone opens more than the
usual estimate for ideal fluid case.

\section{Discussion and Perspectives}

We presented the first full 3D relativistic viscous hydrodynamic calculation
which preserve the causality correctly. In the case of shock wave
simulation, it is fundamental to include the bulk viscosity to take account
for the entropy generation through the shock front. To satisfy the causal
propagation of the signal, the presence of viscosity requires a finite
relaxation time. It is found that the Mach cone angle depends on the
viscosity and the relaxation time. It is also observed that a very turbulent
structure after the shock. More quantitative and systematic study of the
shock wave generation and its propagation in an expanding QGP is in progress.

This work has supported by CNPq, FINEP, FAPERJ and CAPES. T.K. appreciates
the discussion and encouragements of H. Stoecker, E. Fraga and A. Muronga.

\end{document}